\documentclass[prb,floats,twocolumn]{revtex4-2}
\usepackage[english]{babel}
\usepackage{amsmath}
\usepackage{graphicx}

\newcommand{\ntd}{n_\mathrm{2D}}

\begin{document}
\title{Evolution of domain structure with electron doping in ferroelectric thin films}
\date{\today}
\author{W. A. Atkinson} \email{billatkinson@trentu.ca}
\affiliation{Department of Physics and Astronomy, Trent University, Peterborough, Ontario K9L 0G2, Canada}
\begin{abstract}
	To minimize their electrostatic energy, insulating ferroelectric films tend to break up into nanoscale ``Kittel'' domains of opposite polarization that are separated by uncharged 180$^\circ$ domain walls.  Here, I report on self-consistent solutions of coupled Landau-Ginzburg-Devonshire and Schr\"odinger equations for an electron-doped ferroelectric thin film.  The model is based on LaAlO$_3$/SrTiO$_3$ interfaces in which the SrTiO$_3$ substrate is made ferroelectric by cation substitution or strain.  I find that electron doping destabilizes the Kittel domains.  As the two-dimensional electron density $\ntd$ increases, there is a smooth crossover to a zigzag domain wall configuration.  The domain wall is positively charged, but is compensated by the electron gas, which attaches itself to the domain wall and screens depolarizing fields. The domain wall approaches a flat head-to-head configuration in the limit of perfect screening.  The polarization profile may be manipulated by an external bias voltage and the electron gas may be switched between surfaces of the ferroelectric film.
\end{abstract}

\maketitle

\section{Introduction}
\label{sec:introduction}

Interfaces between SrTiO$_3$ and certain polar cap layers, most commonly LaAlO$_3$, become conducting when the cap thickness exceeds a few monolayers \cite{Ohtomo:2004hm,Thiel:2006eo}.  This is primarily due to an electron transfer from the LaAlO$_3$ surface to the SrTiO$_3$ side of the interface \cite{Bristowe:2014fc}, and the resultant electron gas is bound to within a few nm of the interface by residual positive charges on the LaAlO$_3$ surface \cite{Gariglio:2015jx}.  A number of interesting phases, including ferromagnetism \cite{Dikin:2011gl,Bert:2011cg,Kalisky:2012uh}, superconductivity \cite{Reyren:2007gv,Bert:2012el,Richter:2013gn,Klimin:2014ja} and a possible nematic phase \cite{Miao:2016hr,Davis:2017,rout:2017,Boudjada:2018} have been observed.   An attractive feature of these interfaces is that they are tunable:  both the electron doping and spatial profile of the electron density can be modulated by gating \cite{Thiel:2006eo}, and for (001) interfaces there is a narrow doping range near a Lifshitz transition \cite{Joshua:2012bl,Smink:2017cc,Smink:2018tu,Raslan:2018} over which the superconducting transition temperature\cite{Caviglia:2008uh,Dikin:2011gl,Biscaras:2012vd,Maniv:2015cc,Hurand:2015cf}, superfluid density \cite{Bert:2012el}, spin-orbit coupling \cite{Hurand:2015cf,Caviglia:2010jv,BenShalom:2010kv}, and the metamagnetic response \cite{Joshua:2013wl} change by an order of magnitude.

In part, this tunability is due to SrTiO$_3$'s dielectric properties.
SrTiO$_3$ lies close to a quantum critical point separating ferroelectric and paraelectric phases \cite{Muller:1979wa},  
and its dielectric function therefore depends strongly on both temperature and electric field \cite{Hemberger:1995dd}.  This has a profound effect on the interfacial band structure \cite{Copie:2009,stengel:2011,Khalsa:2013hk,Raslan:2017gh,Atkinson:2017jt}.  
Of particular interest for this work, SrTiO$_3$ may be made ferroelectric by cation substitution, as with Sr$_{1-x}$Ca$_x$TiO$_3$ \cite{bednorz84} and Sr$_{1-x}$Ba$_x$TiO$_3$ \cite{lemanov:1996}, by oxygen isotope substitution \cite{Itoh:1999eq}, and by application of lattice strains \cite{Uwe:1976fj,Haeni:2004gj}.  This naturally introduces another tunable parameter---the ferroelectric polarization---with which one may control the electron gas.

Steps in this direction have been taken by several groups in recent years.  Zhou {\em et al.} \cite{zhou_artificial_2019} observed that a 2D electron gas coexists with a ferroelectric-like lattice polarization in LaAlO$_3$/Sr$_{0.8}$Ba$_{0.2}$TiO$_3$ interfaces;  Br\'ehin {\em et al.} \cite{brehin_switchable_2020} demonstrated that Al/Sr$_{0.99}$Ca$_{0.01}$TiO$_3$ interfaces are metallic and switchable; and Tuvia {\em et al.} \cite{Tuvia:2020} obtained hysteretic polarization and resistivity at LaAlO$_3$/Sr$_{1-x}$Ca$_x$TiO$_3$ interfaces.  These experiments raise at least two important questions.

First, why is the polarization switchable?  There were, until recently, reasonable expectations that the itinerant electrons would screen external electric fields and effectively eliminate the ability to manipulate the lattice polarization.  Why does this not happen?

Second, what do the polarization and electron gas profiles look like?  In paraelectric LaAlO$_3$/SrTiO$_3$ interfaces, the electron gas forms a compact 2D layer adjacent to the interface while the lattice polarization points perpendicular to the interface, into the substrate \cite{Gariglio:2015jx}.  On the other hand, it is a universal feature of ferroelectrics that they break up into domains to reduce the large electrostatic depolarizing fields that accompany ferroelectricity \cite{kittel:1946,bennett:2020}. Which, if either, scenario applies to ferroelectric LaAlO$_3$/Sr$_{0.99}$Ca$_{0.01}$TiO$_3$ interfaces?  If domains do form, does the electron gas attach itself to the domain walls, remain at the interface, or do something else?

\begin{figure}[t]
	\includegraphics[width=\columnwidth]{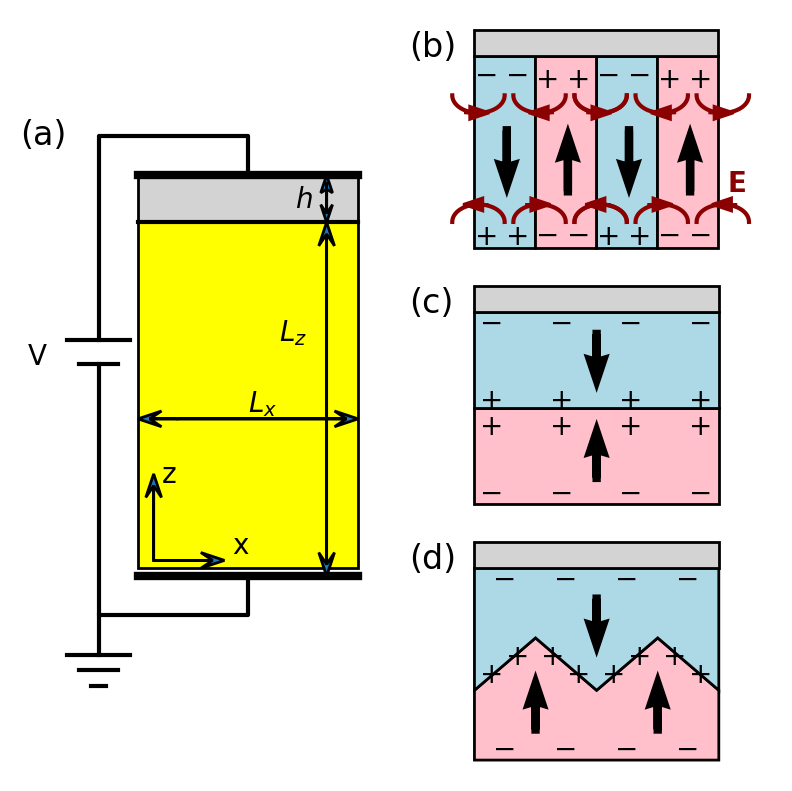}
	\caption{Typical ferroelectric device and representative domain structures.  (a)  Schematic diagram showing a device consisting of a ferroelectric layer of thickness $L_z$ and a dielectric cap layer of thickness $h$ sandwiched between capacitor plates at voltage bias $V$.    The system has periodic boundary conditions in the $x$- and $y$-directions with periodicities $L_x$ and $L_y$, respectively.  The positive $y$ axis points into the page. Representative Kittel (b), head-to-head (c), and zigzag (d) domain structures.  Polarization directions are indicated by thick black arrows.  As shown in (b), the depolarizing fields for the Kittel domains are restricted to the ferroelectric surface region.  The fields in (c) and (d) (not shown) span the whole film, unless they are screened by an electron gas.  }
	\label{fig:Intro}
\end{figure}

K.~Chapman and I recently explored these questions with a model system similar to the one shown in Fig.~\ref{fig:Intro}(a) \cite{chapman:2022}.  The system comprises a ferroelectric substrate capped by a thin dielectric film (representing LaAlO$_3$) and sandwiched by capacitor plates on the top and bottom surfaces.  The cap layer is insulating, and a free electron gas resides in the ferroelectric substrate.  From this model, we obtained a simple explanation for the switchability:  we showed that unless the external field is large, the electron gas attaches itself to polarization gradients and is therefore unavailable to screen external fields.  We further obtained robust hysteresis for the polarization versus bias voltage, and a novel low-polarization branch with negative dielectric susceptibility.  These calculations suggest that when the electron density is comparable to the polarization the electron gas strongly influences the ferroelectric properties. However, these results are based on explicitly one-dimensional (1D) solutions for the polarization, electron density, and electrostatic potential, which are therefore translationally invariant parallel to the interface.

It is known that under most conditions insulating ferroelectrics spontaneously break translational symmetry parallel to the interface and form Kittel domains \cite{kittel:1946,bennett:2020} as a way of reducing the depolarizing electric fields generated by the lattice polarization [Fig.~\ref{fig:Intro}(b)].  In this limit, our previous solutions are incorrect.  On the other hand, we showed that in the limit of large electron densities, these depolarizing fields are screened so that the arguments leading to Fig.~\ref{fig:Intro}(b) break down.  We argued that domain structures like the one shown in Fig.~\ref{fig:Intro}(c) might be stable instead, and indeed Ref.~\cite{Tuvia:2020} reported that the polarization near their LaAlO$_3$/Sr$_{1-x}$Ca$_x$TiO$_3$ interfaces was everywhere into the substrate.  This raises the question: is the Kittel domain structure relevant to electron-doped ferroelectrics.?

To address this, I report on a set of numerical calculations similar to those we published in Ref.~\cite{chapman:2022}, but with the potential, polarization, and electron density allowed to be functions of two spatial coordinates [$x$ and $z$ in Fig.~\ref{fig:Intro}(a)].  I find that the domain structures shown in Fig.~\ref{fig:Intro} are endpoints, with insulating ferroelectric films having domains like those in Fig.~\ref{fig:Intro}(b) and metallic films approaching Fig.~\ref{fig:Intro}(c) in the limit of strong screening of depolarizing fields by the electron gas. At electron densities between the two endpoints, the domains pass from a Kittel-like structure to a zigzag configuration [Fig.~\ref{fig:Intro}(d)].  Although this zigzag configuration is quantitatively different from the 1D solution shown in Fig.~\ref{fig:Intro}(c), the two are qualitatively similar.

The model and calculations are described in Sec.~\ref{sec:methods}.  In brief, the approach is to simultaneously solve Landau-Ginzburg-Devonshire (LGD) and Schr\"odinger equations for the polarization and electron density for the device pictured in Fig.~\ref{fig:Intro}(a).  These equations are coupled via Gauss' law for the electrostatic field, which appears in both equations.  The results of these calculations are shown in Sec.~\ref{sec:results}, and are then discussed in the context of earlier work in Sec.~\ref{sec:discussion}.

\section{Methods}
\label{sec:methods}

Figure~\ref{fig:Intro}(a) shows the device to be modeled.  It consists of a dielectric cap layer of thickness $h$ on top of a ferroelecric substrate of thickness $L_z$.  The system is periodic in the $x$ and $y$ directions with periodicity $L_x$ and $L_y$, respectively, and has translational invariance along the $y$ axis (which points into the page). The dielectric/ferroelectric bilayer is sandwiched by capacitor plates on the top and bottom surfaces, and these are maintained at a potential difference $V$.

In this model, there is a free electron gas in the ferroelectric substrate and the dielectric cap layer is insulating.  The electron gas is assumed to come from a charge transfer between the surface of the cap layer and the substrate, as occurs in the LaAlO$_3$/SrTiO$_3$ system.  Overall charge neutrality is implicit in the assumption that the electric field vanishes outside the capacitor plates.

The total energy has three parts,
\begin{equation}
\tilde {\cal F}[{\bf P}^\mathrm{tot}, \rho^f, \rho^\mathrm{ext}] = {\cal F}_p + {\cal F}_e + {\cal V}
\end{equation}
where ${\cal F}_{p}$ and ${\cal F}_e$ are the free energies due to the polarization and electronic degrees of freedom, respectively.  These exclude contributions from the electric field, which are collected together in the electrostatic energy ${\cal V}$.  As written, the free energy depends on the total lattice polarization ${\bf P}^\mathrm{tot}$, the itinerant charge density due to the electron gas, $\rho^f$, and the external charge density on the capacitor plates and cap-layer surface, $\rho^\mathrm{ext}$.  

It is more meaningful to write the free energy in terms of the  potential $V$ than the external charge density.  This is done via a Legendre transformation,
\begin{equation}
 {\cal F}[{\bf P}^\mathrm{tot}, \rho^f, V] = \tilde {\cal F}[{\bf P}^\mathrm{tot}, \rho^f, \rho^\mathrm{ext}]  - \oint \phi_S \sigma^\mathrm{ext} da,
 \label{eq:F}
\end{equation} 
where $\phi_S$ is the potential on the bounding surface (the capacitor plates) and the external charge  is written as a surface charge density.  For simplicity, the bottom plate in Fig.~\ref{fig:Overview} is grounded ($\phi_S = 0$) and the top plate is at potential $\phi_S=V$.  Equation~(\ref{eq:F}) therefore simplifies to ${\cal F} = \tilde {\cal F} - \sigma^\mathrm{ext} V L_x L_y$.  Note that the Legendre transformation does not change the self-consistent equations for ${\bf P}^\mathrm{tot}$ and $\sigma^f$, but is necessary when the free energies of two solutions at fixed $V$ are to be compared.  In this case $\sigma^\mathrm{ext}$ can be obtained from the electric field at the capacitor plate.

The polarization is assumed to lie in the $x$-$z$ plane, and can be broken up into contributions from the soft ferroelectric phonon mode, ${\bf P}$, and the background polarizability ${\bf P}^b$, such that  ${\bf P}^\mathrm{tot} = {\bf P} + {\bf P}^b$.  The LGD energy for the polarization is then
\begin{widetext} 
\begin{eqnarray}
\frac{{\cal F}_p}{L_y} &=& \int_0^{L_x} dx \int_{0}^{L_z} dz
\left[ a_1 P_x^2 + a_3 P_z^2 + a_{11} [P_x^4 + P_z^4] + a_{12}  P_x^2P_z^2 + 
\frac{g_{11}}{2} \left | \nabla\cdot {\bf P} \right |^2 + \frac{g_{44}}{2} \left[ \left( \frac{\partial P_z}{\partial x} \right )^2  + \left( \frac{\partial P_x}{\partial z} \right )^2\right ] \right ] \nonumber \\
&& +
\int_0^{L_x} dx \int_{0}^{L_z+h} dz \frac{|{\bf P}^b|^2}{2\varepsilon_0\chi(z)}. 
\label{eq:LGDFE}
\end{eqnarray}
\end{widetext}
The LGD parameters  correspond roughly to those found in SrTiO$_3$, and are given in Table~\ref{tab:params}.  A spontaneous polarization along the $z$ axis is induced by taking $a_3<0$; however, $a_1$ is kept positive to avoid trivial solutions in which the ferroelectric spontaneously polarizes along the $x$ axis.  In fact, $a_3<0<a_1$ may be realized by applying a compressive strain along the $x$ and $y$ axes of a SrTiO$_3$ thin film \cite{Haeni:2004gj}.
The background polarization is obtained from the noncritical contribution, $\chi(z)$, to the dielectric susceptibility.  For concreteness, 
\begin{equation}
\chi(z) = \left \{ \begin{array}{lr}
\chi_\mathrm{STO}, & 0 < z < L_z \\
\chi_\mathrm{LAO}, & L_z < z < L_z +  h
\end{array} \right .,
\end{equation}
with $\chi_\mathrm{STO}$ and $\chi_\mathrm{LAO}$ given in Table~\ref{tab:params}.

\begin{table}[tb]
	\begin{tabular}{c|c|c}
		Parameter & Value & Units\\
		\hline
		$a_1$ &  $1.0\times 10^8$ & C$^{-2}$m$^2$N \\
		$a_3$ &  $-8.0\times 10^7$ & C$^{-2}$m$^2$N \\
		$a_{11}$  & $1.70\times 10^9$ & C$^{-4}$m$^6$N \\ 
		$a_{12}$ & $1.47\times 10^9$ & C$^{-4}$m$^6$N \\ 
		$g_{11}$ & $1\times 10^{-10}$ &  C$^{-2}$m$^4$N \\
		$g_{44}$ & $1\times 10^{-10}$ &  C$^{-2}$m$^4$N \\
		$\chi_\mathrm{LAO}$ & 25 & -\\
		$\chi_\mathrm{STO}$ & 4.5 & -\\
		\hline
	\end{tabular}
	\caption{LGD Model Parameters.  Parameters are based on values for SrTiO$_3$ published in Appendix A of Ref.~\cite{Littlewood_review_2017} with the following exceptions:  $a_3$ is negative to so that the model is ferroelectric and gives a bulk polarization $P^\mathrm{bulk}\approx 15$~$\mu$C/cm$^2$;  $g_{11}$ and $g_{12}$ are chosen to give correlation lengths $\xi \approx 1$~nm.  $\chi_\mathrm{LAO}$ is the dielectric susceptibility of LaAlO$_3$, while $\chi_\mathrm{STO}$ is the background susceptibility obtained from the optical dielectric constant for SrTiO$_3$, $\epsilon_\infty \approx 5.5\epsilon_0$ \cite{zollner00}. }
	\label{tab:params}
\end{table}

The itinerant electrons are treated within a single-band effective mass approximation, for which the free energy is
\begin{equation}
{\cal F}_e = -k_BT \ln\left[ e^{-\beta [\hat H - \mu \hat N]} \right ] + \mu N + e\int d{\bf r} \rho^f({\bf r}) \phi({\bf r}).
\label{eq:ElFE}
\end{equation}
Here $\beta = 1/k_BT$, $\rho^f({\bf r})$ is the free electron density, and
\begin{equation}
\hat H = \int d{\bf r}  \hat \Psi^\dagger({\bf r}) 
\left [-\frac{\hbar^2 \nabla^2 }{2m^\ast} - e \phi({\bf r}) \right ] \hat \Psi({\bf r}),
\end{equation}
is the mean-field Hamiltonian with electrostatic potential $\phi({\bf r})$. $\hat \Psi({\bf r})$ is the second-quantized electron annihilation operator.  

The final two terms in Eq.~(\ref{eq:ElFE}) require explanation:  $\mu N$ is a Legendre transformation from the grand potential (fixed chemical potential $\mu$) to the Helmholtz potential (fixed electron number $N$), and is introduced because $\mu$ is adjusted at each step of the calculation to keep $N$ fixed.  The average 2D electron density is related to the electron number by
\begin{equation}
\ntd = \frac{N}{L_xL_y}.
\end{equation}
The final term in Eq.~(\ref{eq:ElFE}) subtracts off the electrostatic energy that is added to the mean-field Hamiltonian $\hat H$; this is to avoid double-counting terms that are included in the electrostatic energy ${\cal V}$.  

Indeed, neither the LGD energy, Eq.~(\ref{eq:LGDFE}), nor the electronic energy, Eq.~(\ref{eq:ElFE}), contains a net contribution from the electric field.  The field energy contributions are collected together in
\begin{equation}
{\cal V} = \frac{1}{8\pi \epsilon_0} \int d{\bf r} \int d{\bf r}' \frac{(\rho^f + \rho^b + \rho^\mathrm{ext})_{\bf r} (\rho^f + \rho^b + \rho^\mathrm{ext})_{\bf r'}}{|{\bf r}-{\bf r'}|}, 
\label{eq:V}
\end{equation}
where
\begin{equation}
\rho^b = -\nabla \cdot {\bf P}^\mathrm{tot} = -\nabla \cdot {\bf P} -\nabla \cdot {\bf P}^b,
\end{equation} 
is the bound charge density.

Minimizing ${\cal F}$ with respect to different components of the polarization leads to the constituent equations,
\begin{eqnarray}
E_z &=& 2a_3P_z + 4a_{11} P_z^3 + 2 a_{12} P_x^2 P_z -g_{11} \partial^2_zP_z - g_{44}  \partial^2_xP_z, \nonumber \\
 \label{eq:Pz} \\
E_x &=& 2a_1P_x + 4a_{11} P_x^3 + 2 a_{12} P_z^2 P_x -g_{11}
\partial^2_x P_x - g_{44} \partial^2_z P_x \nonumber \\
 \\
 {\bf P}^b &=& \epsilon_0 \chi(z) {\bf E}. 
\end{eqnarray}
These are supplemented by the self-consistent equation for the itinerant charge density,
\begin{equation}
\rho^f({\bf r}) = -en_e({\bf r}) = -e \langle \hat \Psi^\dagger({\bf r}) \Psi({\bf r}) \rangle,
\label{eq:rhof} 
\end{equation}
the electrostatic potential
\begin{equation}
-\epsilon_0 \nabla^2 \phi({\bf r}) = \rho^f({\bf r}) + \rho^b({\bf r}),
\label{eq:potential}
\end{equation}
and the boundary conditions
\begin{eqnarray}
{\bf P}(x, 0) &=& 0, \quad \partial_z {\bf P}(x,z)|_{z=L_z} = 0, \\
{\bf P}(x,z) &=& {\bf P}(x+L_x,z), \\
\phi(x,0) &=& 0, \quad \phi(x, L_z+h) = V, \label{eq:bc} \\
\phi(x,z) &=& \phi(x+L_x,z).
\end{eqnarray}
Setting the polarization to zero at $z=0$ physically separates the external charge density on the lower capacitor plate from the bound surface charge density ${\bf P}\cdot (-{\bf \hat z})$ at the lower surface of the ferroelectric; this is convenient for analysis but has no consequences for the main results of this paper.  Note also that Eq.~(\ref{eq:potential}) does not explicitly include $\rho^\mathrm{ext}$; this contribution is implicit in the boundary condition (\ref{eq:bc}). 

Equations~(\ref{eq:Pz})-(\ref{eq:potential})  must be solved self-consistently for the polarization, itinerant charge density, and potential.
It should be noted that evaluation of Eq.~(\ref{eq:rhof}) is by far the slowest step because it requires solving Schr\"odinger's equation for the eigenstates of $\hat H$.  Because materials like Sr$_{1-x}$Ca$_{x}$TiO$_3$ have low transition temperatures ($T_c \lesssim 30$~K), we focus on the $T\rightarrow 0$ limit; however, self-consistency of the Schr\"odinger equation can often be improved by taking a finite value of $T$ that masks artificial discreteness of the electronic band structure due to finite-size effects.  For this reason, $\rho^f$ is evaluated at temperature $T = 10$~K.  Further details are given in the Appendix.

\section{Results}
\label{sec:results}
\subsection{Short-circuit conditions}

\begin{figure}[tb]
	\includegraphics[width=\columnwidth]{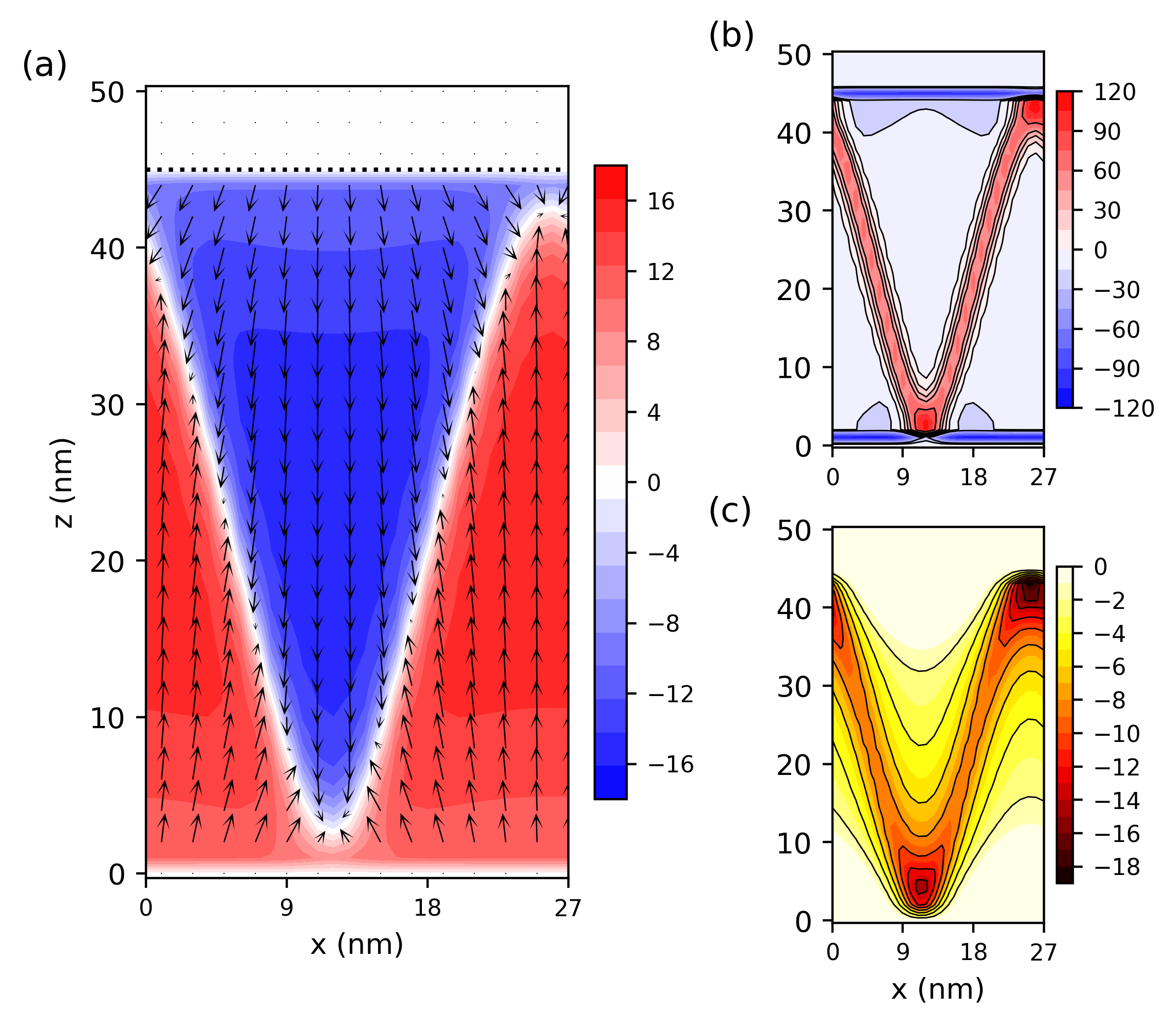}
	\caption{Typical self-consistent solution showing (a) the lattice polarization ${\bf P}^\mathrm{tot}$, (b) the bound charge density $\rho^b = -\nabla \cdot {\bf P}$, and (c)  the free charge density $\rho^f=-en_e$.  Results are for a $28\times 46$~nm ferroelectric substrate with a 5~nm dielectric cap layer.  The average electron density is $\ntd=0.20$ per 2D unit cell and charge densities are in C/cm$^3$.  Arrows in (a) indicate the polarization direction while the color indicates the value of $P_z$ in $\mu$C/cm$^2$.  Results are for the short-circuit boundary condition, $V=0$.  Other parameters are given in Table~\ref{tab:params}.}
	\label{fig:Overview}
\end{figure}

Typical self-consistent results for the polarization and electron density are pictured in Fig.~\ref{fig:Overview} for the short-circuit configuration ($V=0$).  Because the LGD parameters in Eq.~(\ref{eq:LGDFE}) are chosen such that $a_3 < 0 < a_1$, the ferroelectric polarizes spontaneously along the $z$-axis and only tilts towards the $x$-axis near domain walls [Fig.~\ref{fig:Overview}(a)].  The average electron density is chosen to be $\ntd = 0.20$~electrons per 2D unit cell, which corresponds to a 2D charge density $e \ntd = 20$~$\mu$C/cm$^2$ assuming a lattice constant $a=4$~\AA.  This is similar to the maximum polarization in the ferroelectric domains ($P^0 \approx 15$~$\mu$C/cm$^2$), so that the bound and free charges make comparable contributions to the electric field.  The  domain wall structure shown in Fig.~\ref{fig:Overview}(a) is a consequence of feedback between these two contributions. 

The polarization profile in Fig.~\ref{fig:Overview}(a) is striking because it has a zigzag-shaped domain wall separating regions of positive and negative polarization.  This is different from the Kittel domains found in insulating ferroelectric films because there is a substantial positive bound charge density associated with the domain wall, while the top and bottom surfaces of the ferroelectric substrate both have negative  charge densities [Fig.~\ref{fig:Overview}(b)].  The electron gas is attached, albeit loosely, to the domain wall and partially screens the positive bound charge [Fig.~\ref{fig:Overview}(c)].  By construction, the electron density is constant along the $y$ axis (into the page).

\begin{figure*}[tb]
	\includegraphics[width=0.8\textwidth]{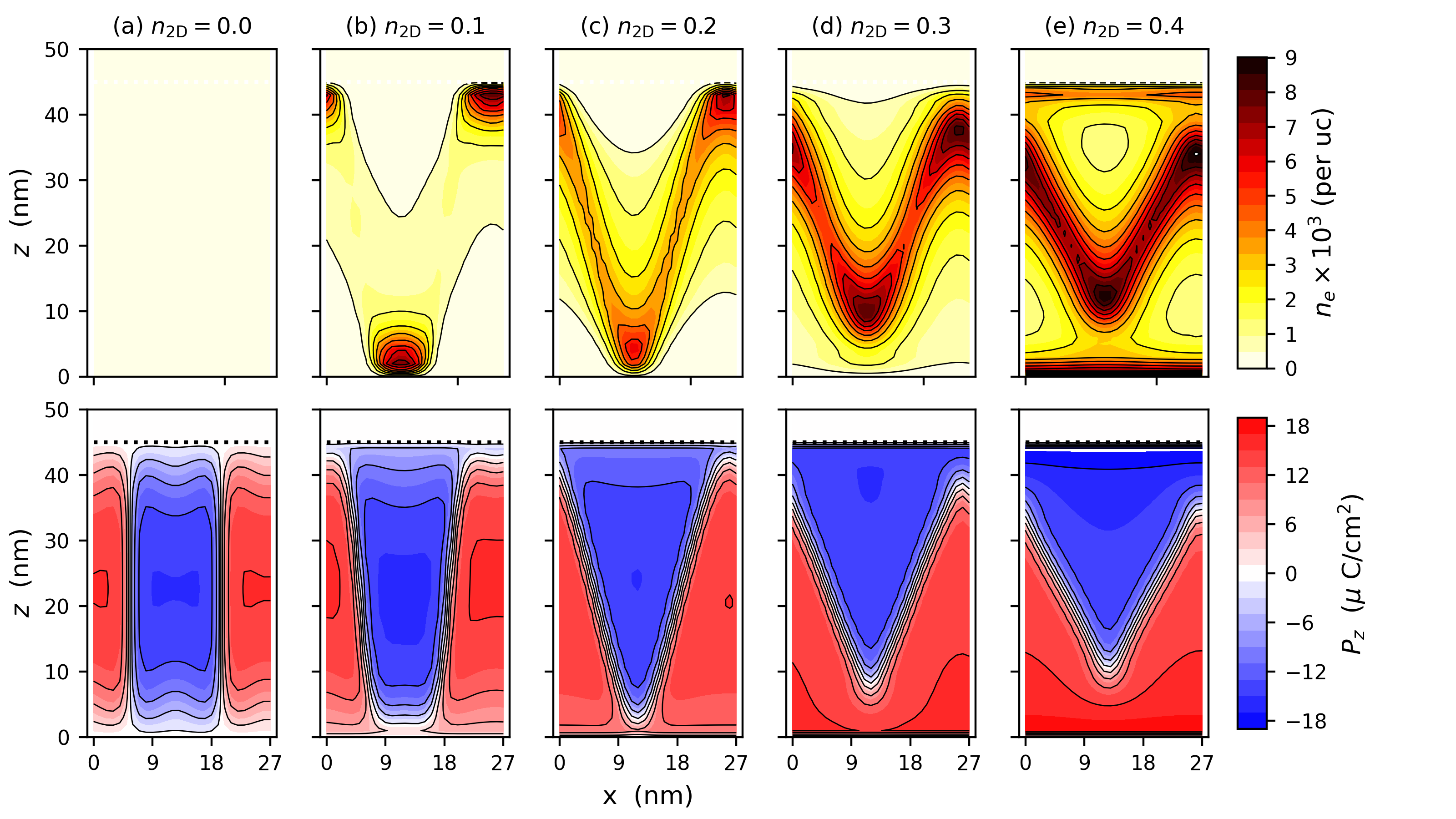}
	\caption{Electron density (top row) and $z$-component of the polarization (bottom row)  in the short-circuit configuration ($V = 0$).  Results are for (a) $\ntd=0.0$, (b) $\ntd=0.1$, (c) $\ntd=0.2$, (d) $\ntd=0.3$, and (e) $\ntd=0.4$ electrons per 2D unit cell.  Other parameters are as in Fig.~\ref{fig:Overview}. }
	\label{fig:polarization}
\end{figure*}

Figure~\ref{fig:polarization} illustrates how the the zigzag profile emerges from Kittel domains  and then evolves with increasing $\ntd$.  The figure shows both the electron density and $z$-component of the polarization for a series of $\ntd$ values spanning the range $e\ntd \ll P^0$ to $e\ntd \gtrsim 2P^0$.    In all cases, the maximum polarization amplitude $P^0$ in the ferroelectric domains differs only slightly from the nominal bulk polarization 
\[
P^\mathrm{bulk} \equiv \sqrt{\frac{-a_{3}}{2a_{11}}},
\]
which is held fixed. 

When $\ntd = 0$ [Fig.~\ref{fig:polarization}(a)], the ferroelectric film breaks up into Kittel domains, namely oppositely polarized domains separated by $180^\circ$ domain walls along which $\rho^b=0$. The bound charge density is nonzero along the top and bottom surfaces of the ferroelectric substrate.  Taking $\rho^b \approx -\partial P_z/\partial z$, one can assign positive bound charges to the top and bottom ends of the red and blue domains, respectively.

For small electron densities [Fig.~\ref{fig:polarization}(b)], the electrostatic energy is dominated by contributions from the lattice polarization. The domain structure is therefore of the Kittel type and the electron gas attaches itself to the positively charged ends of the Kittel domains to form one-dimensional conducting channels along the $y$ axis.  These one-dimensional channels are weakly connected by low density tails, and the resulting electron density profile in Fig.~\ref{fig:polarization}(b) has a zigzag shape in the $x$-$z$ plane.  

Although the domains in Fig.~\ref{fig:polarization}(b) have the Kittel structure  to a first approximation, they have been modified by the electron gas in two significant ways.  First, the positive ends of the domains have moved inward from the top and bottom surfaces of the ferroelectric, relative to Fig.~\ref{fig:polarization}(a). The polarization thus points into the substrate everywhere along the surfaces, which are in consequence negatively charged.  Second, the widths of the positively charged domain ends have shrunk while the negative domain ends have expanded, relative to Fig.~\ref{fig:polarization}(a), with the overall effect that the domain walls have tilted slightly.  This tilting causes the domain walls to develop a small positive charge density, which is partially compensated by the negatively charged electron gas tails that extend along the walls.

The domain wall tilt angle increases with increasing $\ntd$ until the domain walls connect to form a zigzag pattern [Fig.~\ref{fig:polarization}(c)]. The tilt angle continues to increase as $\ntd$ grows further, and the ends of the zigzag domain wall move inwards from the surfaces of the ferroelectric [Fig.~\ref{fig:polarization}(d) and (e)].  
At sufficiently large electron densities [Fig.~\ref{fig:polarization}(e)], a fraction of the electron gas spills over from the domain wall to the interface region.  We observed similar behaviour in Ref.~\cite{chapman:2022} for idealized head-to-head domain walls where, as with Fig.~\ref{fig:polarization}, electrons spilled over to the surfaces when $e\ntd \gtrsim 2P^0$.

\begin{figure}[tb]
	\includegraphics[width=\columnwidth]{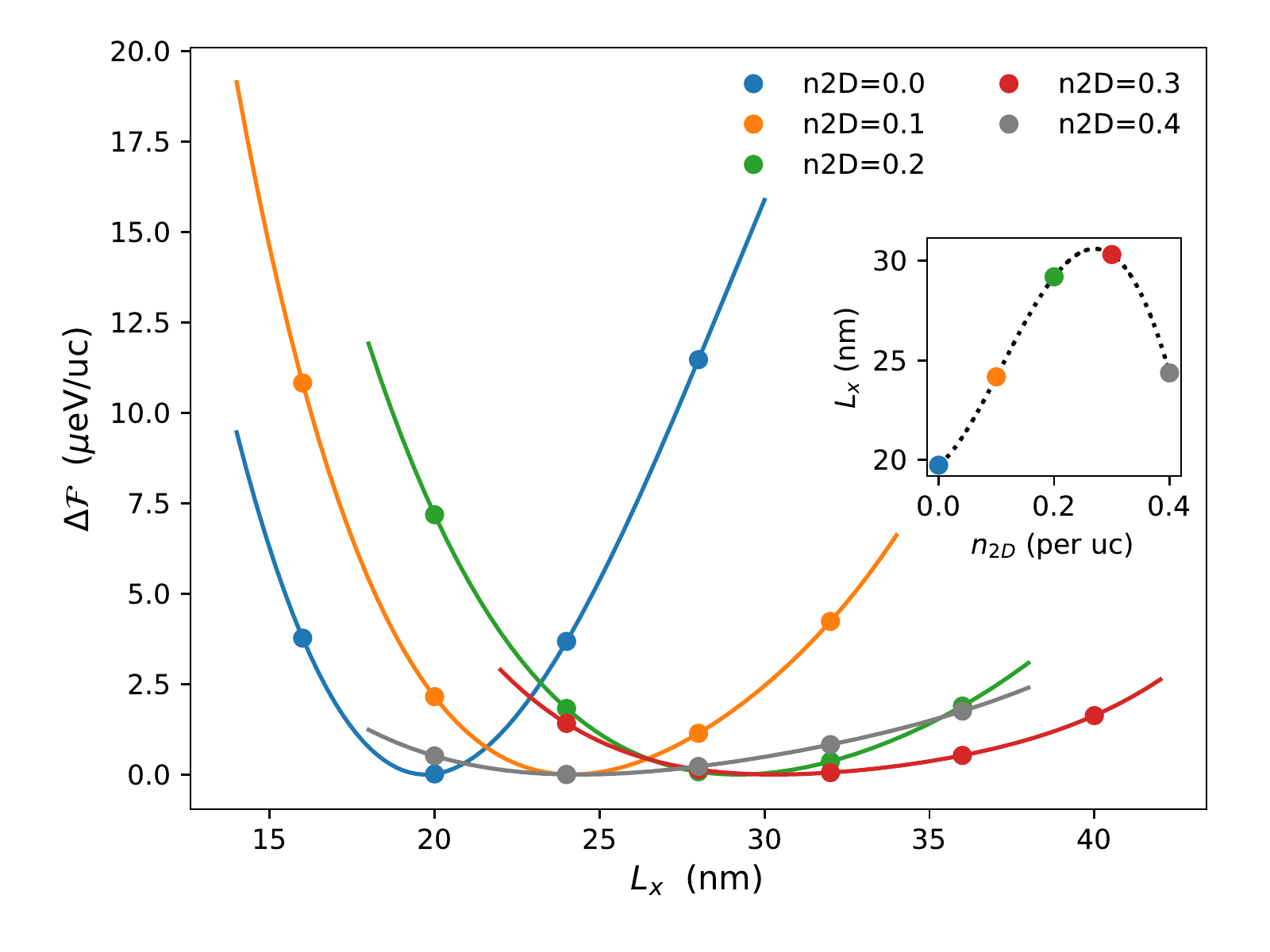}
	\caption{Domain wall periodicity.  The main panel shows the dependence of $\Delta {\cal F} \equiv {\cal F} - {\cal F}_\mathrm{min}$ on $L_x$ for different electron densities.  Solid points are data, lines are polynomial fits to the data.  The inset shows the value of $L_x$ at which the polynomial fits are minimized as a function of $\ntd$.  Results are for $V=0$.}
	\label{fig:DWsize}
\end{figure}

It should be noted that the results shown in Fig.~\ref{fig:polarization} are for a fixed periodicity $L_x$ that does not necessarily minimize the free energy ${\cal F}$.   Figure~\ref{fig:DWsize} shows the energy $\Delta {\cal F} = {\cal F} - \min_{L_x}({\cal F})$ as a function of $L_x$ for five different values of $\ntd$.  The optimal domain-wall periodicity corresponds to the value of $L_x$ at which $\Delta {\cal F}=0$.  The optimal $L_x$ is plotted as a function of $n_\mathrm{2D}$ in the figure inset. Two trends are apparent in Fig.~\ref{fig:polarization}.  First, $\Delta {\cal F}$ has the strongest dependence on $L_x$ when $\ntd=0$, and this becomes progressively weaker as $\ntd$ increases.  This suggests that screening by the electron gas softens the domain walls.
Second,  the optimal periodicity is a nonmonotonic function of $\ntd$:  $L_x$ increases with $\ntd$ when the domain walls have a Kittel-like structure (that is, while there are many distinct domain walls, each connecting the top and bottom surfaces of the ferroelectric), and decreases with $\ntd$ when the domain walls connect to form a single zigzag wall. 

Some aspects of Figs.~\ref{fig:polarization} and \ref{fig:DWsize} can be understood in terms of the competition between electrostatic and domain wall energies \cite{bennett:2020}.  The key point is that electric fields are progressively screened as $\ntd$ increases, such that the relative importance of the domain wall energy increases.  The domain wall energy is proportional to the domain wall area and at low electron densities, where the domain walls are Kittel-like, this is minimized by maximizing the spacing between them.   This naturally explains why $L_x$ increases with $\ntd$.
At high electron densities, where there is a single zigzag wall, the domain-wall area is minimized by increasing the tilt angle towards $90^\circ$ (i.e.\ towards the horizontal).  There, one expects the zigzag domain wall to approach a flat horizontal configuration as $\ntd$ increases.  Both of these trends are seen in the numerics, as discussed above. 

\begin{figure}
\includegraphics[width=\columnwidth]{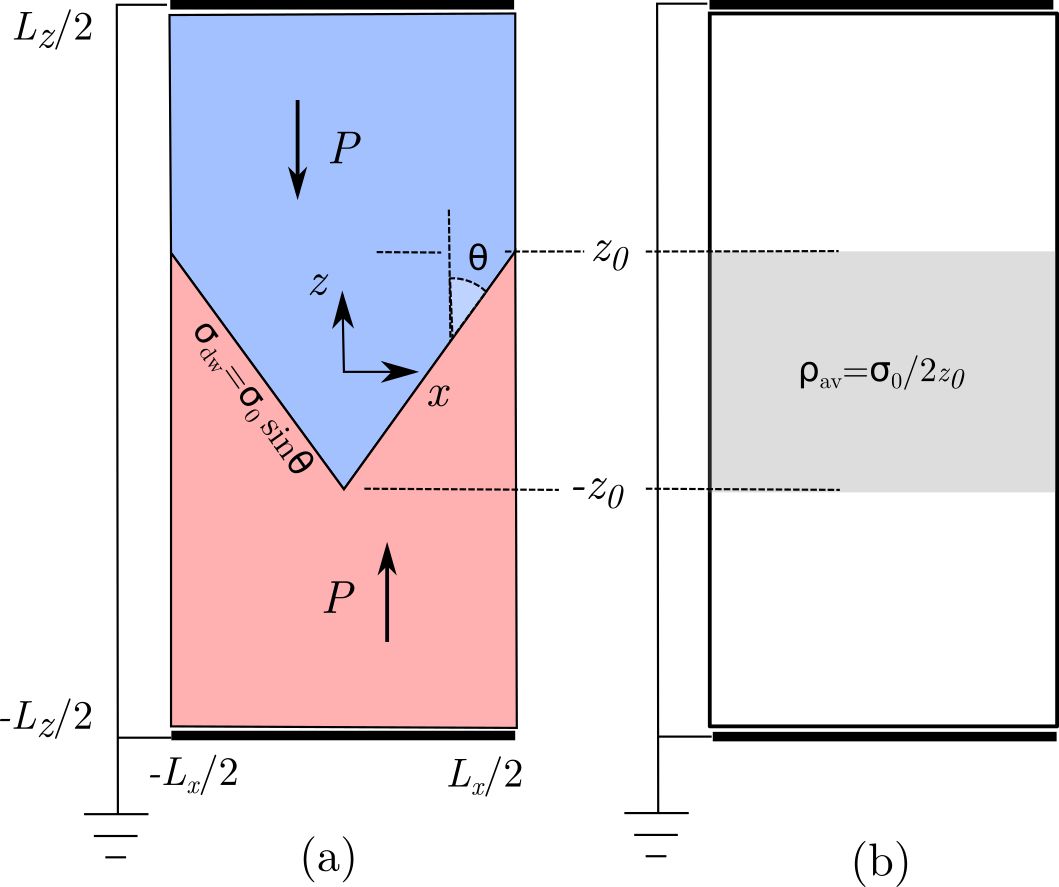}	\caption{Toy model for the energetics of zigzag domain walls.  (a) The model consists of a ferroelectric in the short-circuit configuration with a zigzag domain wall separating regions of opposite polarization. The electron gas is assumed to be uniformly distributed along the domain wall. The domain wall is tilted by an angle $\theta$ and hosts a net charge density $\sigma_\mathrm{dw}$ given by Eq.~(\ref{eq:sigmadw}).  The zigzag pattern extends over the interval $[-z_0, z_0]$ in the $z$ direction.  (b)  As a further simplification, the charge is treated as if it is uniformly spread over $[-z_0, z_0]$ with an average charge density $\rho_\mathrm{av} = \sigma_0/2z_0$.  }
	\label{fig:toymodel}
\end{figure}

The physics of the zigzag domain wall is captured by a simple toy model, pictured in Fig.~\ref{fig:toymodel}. In this model, a ferroelectric is sandwiched between two grounded conducting plates and hosts a zigzag domain wall.  Itinerant electrons are bound to the domain wall, so the average 2D charge density along the wall is
\begin{equation}
\sigma_\mathrm{dw} = \sigma_0 \sin \theta
\label{eq:sigmadw}
\end{equation}
where $\sigma_0 = 2P^0-e\ntd$ and $\theta$ is the tilt angle.  The grounded plates compensate the surface charges, so the only charges in the system lie along the domain wall.  The zigzag wall extends from $-z_0$ to $z_0$, where $2z_0 = L_x \tan \theta$.  To keep things simple, I make the further approximation that the domain wall charge is uniformly smeared over $[-z_0,z_0]$,  so 
\begin{equation}
\rho(z) = \left \{ \begin{array}{lr}
0, & |z| > z_0 \\
\dfrac{\sigma_0}{2z_0}, & |z| < z_0
\end{array} \right..
\end{equation}  
This is obviously a crude approximation; however, it captures the essential feature that the domain wall charge becomes less spread out as the tilt angle increases.

The electrostatic potential vanishes at the top and bottom plates and is
\begin{equation}
\phi(z) = \left \{ \begin{array}{lr}
\dfrac{\sigma_0}{2\epsilon}\left (\dfrac{L_z}2 -|z| \right ), & |z| > z_0 \\
\dfrac{\sigma_0}{4z_0 \epsilon}( z_0L_z - z_0^2 - z^2  ), & |z| < z_0
\end{array} \right..
\end{equation}
In this expression, $\epsilon$ is the dielectric permittivity of the ferroelectric.

Combining the  electrostatic energy, $U_\mathrm{es} = \frac 12 \int \rho \phi$, and the domain wall energy  $U_\mathrm{dw}$, one obtains the energy density
\begin{eqnarray}
f_\mathrm{toy} &=& \frac{1}{L_xL_yL_z}( U_\mathrm{es} + U_\mathrm{dw}) \nonumber \\
&=& \frac{\sigma_0^2}{4\epsilon L_z} \left( \frac{L_z}2- \frac{2z_0}{3}\right ) + 
 \frac{\Sigma\sqrt{L_x^2 + 4z_0^2}}{L_xL_z},
\end{eqnarray}
where $\Sigma$ is the domain wall energy per unit area.

The optimal height $z_0$ of the zigzag pattern is then obtained by minimizing $f_\mathrm{toy}$ with respect to $z_0$.  From this, one obtains the simple result for the tilt angle,
\begin{equation}
\cos\theta = \frac{L_x (2P^0-e\ntd)^2}{12\epsilon \Sigma}.
\label{eq:tiltangle}
\end{equation}
This equation predicts that the tilt angle increases when $\ntd$ increases, up to a maximum of $\theta = 90^\circ$ when $en_\mathrm{2D} = 2P^0$. (Note that this model only applies to the regime $e\ntd \leq 2P^0$; at higher densities one must allow  electrons to spill over to the surfaces.)  This is not observed in numerical calculations because the electron kinetic energy (not included in the toy model) ensures that electrons spread away from the domain wall and screen internal fields inefficiently.  I have checked numerically that when $m^\ast$ is increased, the electron gas is bound more tightly to the domain wall, and the tilt angle increases as one would expect.

So far, the model provides no insight into what determines the optimal zigzag periodicity.  Indeed, setting $\partial f_\mathrm{toy}/\partial L_x =0$ yields $L_x \rightarrow \infty$.  I have explored more complicated versions of the model, which preserve the zigzag structure of the charge density, but obtain the same qualitative results.  It seems likely that a key missing ingredient is that the electron gas distribution depends nontrivially on the domain wall configuration.  The numerical calculations in Fig.~\ref{fig:polarization} show that the electron density tends to be highest at the vertices of the zigzag domain walls.  This is because there is a shallow potential well at each vertex.  It is plausible that the energy of the system can be lowered by having sufficient vertices that all electrons can be accommodated by low-energy states within these wells.  In this case, the number of vertices should scale with $\ntd$ such that $L_x \propto \ntd^{-1}$.  Qualitatively, this matches the trend shown in Fig.~\ref{fig:DWsize}.  If this explanation is correct, then it indicates that while the tilt angle is determined primarily by the difference between electrostatic and domain wall energies, predicting the zigzag periodicty requires knowledge of the band structure of the electron gas.

\subsection{Voltage Dependence}
\begin{figure*}
	\includegraphics[width=0.8\textwidth]{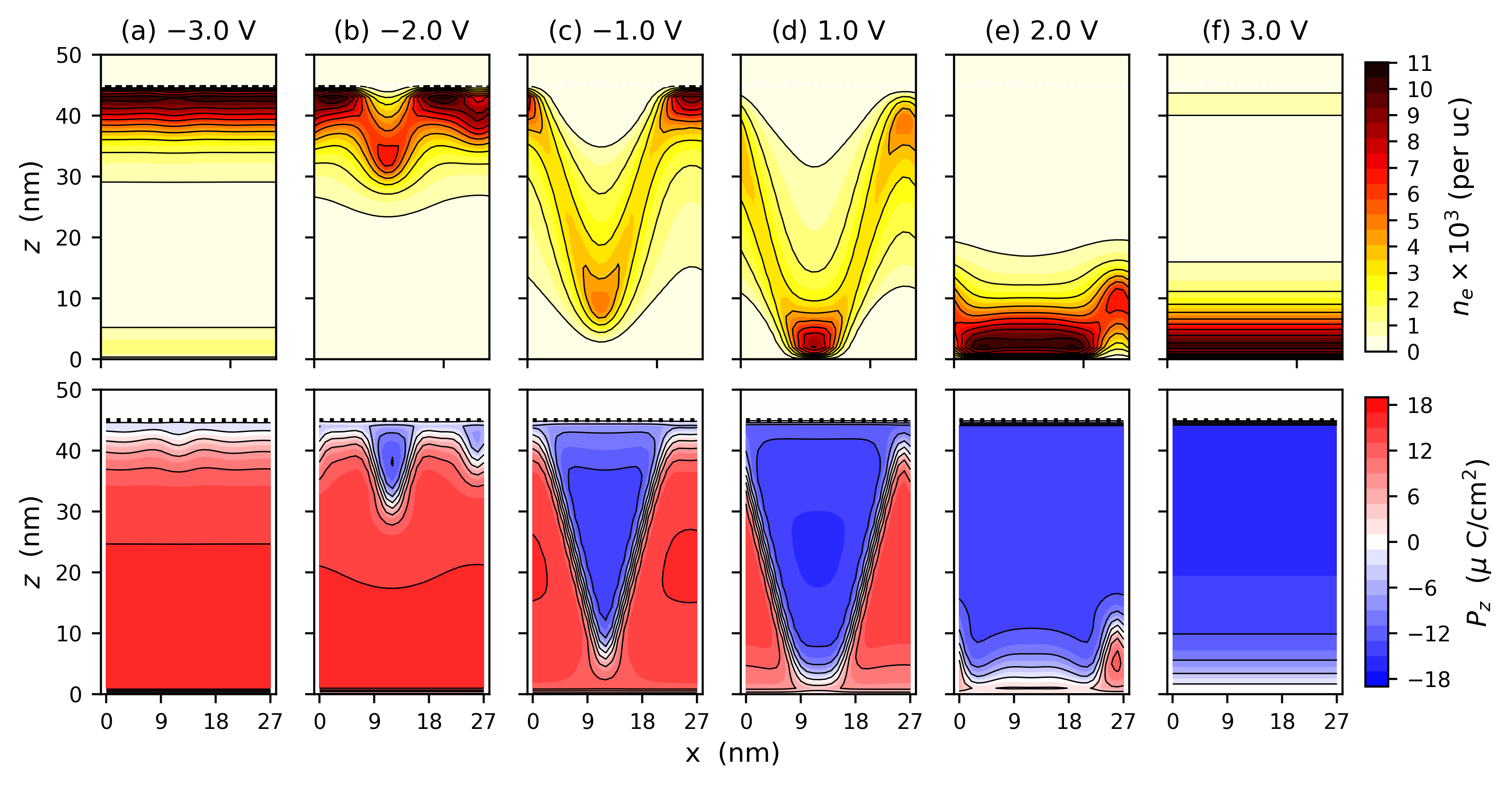}
	\caption{Voltage dependence of the electron density (top row) and polarization (bottom row) for (a) $V=-3.0$~V, (b) $V=-2.0$~V, (c) $V=-1.0$~V, (d) $V=1.0$~V, (e) $V=2.0$~V, and (f) $V=3.0$~V.   Results are for $\ntd=0.20$.  Other parameters are as in Fig.~\ref{fig:Overview}.}
	\label{fig:Vdep}
\end{figure*}

Figure \ref{fig:Vdep} shows the dependence of the electron density and polarization profiles as a function of bias voltage for a fixed $\ntd=0.20$.  By convention, a positive voltage indicates that the top surface is at a higher potential than the bottom surface.

At large negative bias [Fig.~\ref{fig:Vdep}(a)], the polarization points upwards, similar to what one would find in an insulating ferroelectric.  The itinerant electrons are then acted on by two distinct forces:  the external field due to the capacitor plates pushes the electron gas towards the bottom surface, and the internal field due to the polarization gradients draws the electron gas towards the interface.  As we discussed at length in Ref.~\cite{chapman:2022}, it is the internal field that largely controls the behaviour of the itinerant electrons: bound and free charges bind together to form an approximately neutral compensated state at the positive end of the ferroelectric.  The orientation of the polarization is  then controlled by the external field through its action on the unscreened negative end of the ferroelectric.  This scenario begins to break down when $V$ exceeds a threshold ``spillover'' voltage, beyond which some of the electron gas spills over to the negative end of the ferroelectric.  This occurs in Fig.~\ref{fig:Vdep}(a), where a small fraction of the electron gas lives at the bottom of the substrate.

As $V$ is increased to $-2.0$~V, ripples in the polarization profile grow into fingers of opposite polarization that extend into the ferroelectric [Fig.~\ref{fig:Vdep}(b)].  These further evolve into zigzag patterns when $V=-1.0$~V [Fig.~\ref{fig:Vdep}(c)].  The polarization and electron density patterns then reverse themselves at positive bias, with the system obtaining a uniformly polarized state at $V=3.0$~V [Figs.~\ref{fig:Vdep}(d)-(f)].  In Fig.~\ref{fig:Vdep}(f), $V$ exceeds the spillover threshold, and some of the electron gas is bound to the interface. 

A few comments can be made about the sequence shown in Fig.~\ref{fig:Vdep}.  First, although the average polarization changes sign as the voltage is swept from negative to positive (as one might expect), the polarization at the interface is negative for all $V$.  That is, the electron gas actually generates a sign reversal in the polarization at the interface when $V<0$. We observed similar effects in Ref.~\cite{chapman:2022}, and these appear tied to overscreening by the electron gas when $e\ntd \gtrsim P^0$.  
	
Second, it should be emphasized that the results presented in Fig.~\ref{fig:Vdep} are for the lowest energy solutions of the self-consistent equations.  Metastable solutions have been found, but are difficult to stabilize using the numerical schemes adopted in this work.  The solution shown in Fig.~\ref{fig:Vdep}(a), for example, is metastable over a small range of $V > -3.0$~V.

Third, the dielectric susceptibility obtained from the spatially averaged polarization, $\chi = \epsilon_0^{-1} \partial P_\mathrm{av}/\partial E_\mathrm{av}$, is positive.  That is, the average electric field and polarization point in the same direction.  This is different from what we observed for the head-to-head configuration in Ref.~\cite{chapman:2022}, which exhibited negative susceptibility.  The reason for this difference is unclear.

\section{Discussion}
\label{sec:discussion}

Many of the results reported in this paper are consistent with the 1D calculations of Ref.~\cite{chapman:2022}.  There, we observed high-polarization states much like those shown in Fig.~\ref{fig:Vdep}(a) and (f). We found that the electron gas compensates bound charges associated with polarization gradients near the surfaces, so that these regions are close to electrically neutral.  The consequence is that depolarizing fields may be nearly eliminated, but that external fields are largely unscreened.  This result was key to understanding the switchability of the polarization by external fields.  The compensation was found to break down either when $e\ntd > 2P^0$ or when the bias voltage $V$ is greater than a spillover voltage.  In both cases, some fraction of the electron gas remains unattached to bound charges and is free to screen external fields. All of these results are consistent with the 2D calculations described here.  However, there are two important differences between the current and previous calculations.

First, as mentioned above, there is no low-polarization branch with negative dielectric susceptibility in the current work.  In Ref.~\cite{chapman:2022}, this branch was connected to the formation of a head-to-head domain wall running parallel to the film surfaces.  Here, it has been found that the situation is more complicated, namely that $180^\circ$ domain walls in the insulating ferroelectric evolve towards a flat head-to-head domain wall as depolarizing fields are increasingly screened by itinerant electrons.  While the zigzag domain walls are qualitatively similar to the flat head-to-head domain walls reported previously, the dielectric susceptibility that one may infer from Fig.~\ref{fig:Vdep} is positive.

Second, the pronounced hysteresis curves reported in Ref.~\cite{chapman:2022} have not been found in the 2D calculations. Hysteretic tendencies are observed in the numerics (that is, self-consistent calculations appear to converge towards a metastable state before settling on the ground state) and metastable states can be stabilized at large bias voltages, but (for example) high polarization states with the electron gas confined to a single surface have not been found at $V=0$.  It is expected that calculations at finite driving frequencies would find hysteresis curves.

While these differences may simply reflect the additional degrees of freedom inherent in the 2D solution, it is worth noting that the electronic Hamiltonians used here and in Ref.~\cite{chapman:2022} are not the same.  In Ref.~\cite{chapman:2022}, we had three anisotropic bands derived from $t_{2g}$ orbitals of the Ti atoms, while here I have taken a simpler model with a single isotropic band with effective mass $m^\ast$ equal to the bare electron mass.  This distinction is potentially important as the $d_{xy}$-derived bands in the $t_{2g}$ Hamiltonian are heavy along the $z$ direction and are especially effective at screening electric fields.  Further work is needed to understand the influence of the $t_{2g}$ bands on the domain wall structure.

It is instructive to compare the current calculations with recent experiments on LaAlO$_3$/Sr$_{0.99}$Ca$_{0.01}$TiO$_3$ interfaces reported by Tuvia {\em et al.} \cite{Tuvia:2020}.  There, the authors concluded that the ferroelectric Sr$_{0.99}$Ca$_{0.01}$TiO$_3$ substrate was uniformly polarized at the interface, with no evidence of a Kittel-like domain structure.  Furthermore, the authors inferred from their experiments that the ferroelectric polarization creates a depletion layer such that the electron gas moves away from the interface into the substrate.  From this, and from details of the domain patterns of the octahedral tilts, the authors further concluded that the polarization at the interface points into the substrate, along the $z$ direction.  

To make a meaningful comparison, it must be noted that an ungated sample in the experiments corresponds to a positive bias voltage in my calculations; this is because the  LaAlO$_3$ surface has a residual positive charge that is equal and opposite to the free electron charge density.  Furthermore, typical polarizations in bulk Sr$_{0.99}$Ca$_{0.01}$TiO$_3$ are $P^0 \sim 1$-3~$\mu$C/cm$^{-2}$ \cite{Rischau:2017vj}, which is comparable to the electron density that one can infer from the Hall number given in the Supplemental Information of Ref.~\cite{Tuvia:2020}.  Although these values are approximately an order of magnitude smaller than the ones used in the calculations presented here, they correspond to the regime  $e\ntd \sim P^0$ where zigzag domains are preferred over Kittel domains.

With this in mind, the following scenario provides a plausible explanation for the experiments of Ref.~\cite{Tuvia:2020}.  In the ferroelectric phase, a zigzag domain wall forms within the Sr$_{0.99}$Ca$_{0.01}$TiO$_3$ substrate, such that the polarization points inwards at the top and bottom surfaces of the substrate. Because of the comparitively low value of $P^0$, the bound charge densities are small and electrons are only loosely attached to the domain wall.  Importantly, a significant fraction of them spill over to the interface because of the positive charge on the LaAlO$_3$ surface.  The domain wall acts as a charge reservoir, and the fraction of the electrons at the interface can be manipulated by an applied gate voltage.

Finally, I note that the strongest prediction of the model presented here is that the resistivity should be anisotropic, with the anisotropy largest at low electron densities. In this case, one expects the conductivity to be large along the $y$ direction but small in the $x$ direction. Because there are two possible orientations for the zigzag walls (that is, the zigzag may lie in the $x$-$z$ or $y$-$z$ planes), the domain wall patterns will be twinned.  The conductivity anisotropy could then be revealed by detwinning through the application of a uniaxial strain.

\section{Conclusions}

I have explored 2D solutions of coupled Landau-Ginzburg-Devonshire and Schr\"odinger equations to understand the effect of electron doping on thin ferroelectric films.  The results reported here support many of the conclusions that Chapman and I obtained previously  in Ref.~\cite{chapman:2022}.  Namely, I found that the electron gas tends to bind to positively charged polarization gradients to form a compensated state.  Because of this, the electron gas responds only weakly to an applied external field which, in consequence, may be used to manipulate the orientation of the polarization.  The compensated state breaks down either when the bias voltage exceeds a spillover threshold or when the electron density is larger than is needed to completely compensate the bound charge density.  In either case, a fraction of the electron gas is available to partially screen external fields.

There are also some significant differences with Ref.~\cite{chapman:2022}, which are apparent when $V=0$.  The most significant of these is that in Ref.~\cite{chapman:2022} we obtained a horizontal head-to-head domain wall with negative dielectric susceptibility, whereas here I found that Kittel domains evolve into a zigzag domain structure with positive dielectric susceptibility. There are qualitative similarities between the solutions:  Like the horizontal domain wall, the zigzag domain wall has a head-to-head orientation of the polarizations, making it positively charged; in both cases, domain wall formation is enabled by screening of depolarizing fields by the electron gas.

The important conclusion of this work is that the domain structure in ferroelectric films is fundamentally altered by the presence of an electron gas.  An interesting consequence of this is that by removing electrostatic forces as the primary driver of domain formation, one opens the possibility that other (short-range) forces may play a significant role in shaping the ferroelectric state.

\begin{acknowledgments}
	This work is supported by the Natural Sciences and Engineering Research Council (NSERC) of Canada.  
\end{acknowledgments}

\appendix
\section{Numerical Solution of the Schr\"odinger Equation}
The electron gas is modeled by a free electron model with an effective mass $m^\ast$.  The Schr\"odinger equation is
\begin{equation}
\left[ -\frac{\hbar^2}{2m^\ast}\nabla^2 - e\phi(x,z)\right ] \Psi(x,y,z) = \epsilon \Psi(x,y,z).
\end{equation}
Because the potential $\phi$ depends only on $x$ and $z$, we can use separation of variables to write
\begin{eqnarray}
\Psi(x,y,z) &=& \frac{1}{\sqrt{L_y}} e^{ik_yy} Z^{(n)}(x,z), \\
\epsilon_{n,k_y} &=& \tilde \epsilon_{n} + \frac{\hbar^2}{2m^\ast}k_y^2,
\end{eqnarray}
where $\epsilon_{n,k_y}$ is the energy for band $n$ with wavevector $k_y$, and $\tilde \epsilon_n$ is the eigenvalue for the equation
\begin{equation}
-\frac{\hbar^2}{2m^\ast}\left( \frac{\partial^2 Z^{(n)}}{\partial x^2} 
+ \frac{\partial^2Z^{(n)}}{\partial z^2}  \right) - e\phi(x,z) Z^{(n)} = \epsilon_n Z^{(n)}
\label{eq:Schro}
\end{equation}
subject to the boundary conditions
\begin{equation}
Z^{(n)}(x,0) = Z^{(n)}(x,L_z) = 0, \quad Z^{(n)}(0,z) = Z^{(n)}(L_x, z)
\end{equation}
The eigenvalue equation is solved on a grid.  Let the grid contain $n_x\times n_z$ points with index $(i,j) \in [1,n_x]\times[1, n_z]$.  The grid points are spaced by $\Delta$, so $z_j = (j-1)\Delta$.  If the first grid point is at $z=0$ and the final point at $z=L_z$, then 
\begin{equation}
\Delta = \frac{L_z}{n_z-1} = \frac{L_x}{n_x}.
\end{equation}
The discrete approximation for the 2nd derivatives is
\begin{eqnarray}
\frac{\partial^2 Z^{(n)}}{\partial x^2 }  &\approx& \frac{ Z^{(n)}_{i-1,j} + Z^{(n)}_{i+1,j} - 2Z^{(n)}_{i,j} }{\Delta^2}, \\
\frac{\partial^2 Z^{(n)}}{\partial z^2 }  &\approx& \frac{ Z^{(n)}_{i,j-1} + Z^{(n)}_{i,j+1} - 2Z^{(n)}_{i,j} }{\Delta^2},
\end{eqnarray}
with $Z^{(n)}_{1,j} = Z^{(n)}_{n_z,j} =0$ and $Z^{(n)}_{n_x+1,j} = Z^{(n)}_{1,j}$.
Equation~(\ref{eq:Schro}) then becomes the eigenvalue problem
\begin{equation}
\left [ \begin{array}{ccccc} 
h_{2} & b_\perp & 0 & \ldots &0 \\
b_\perp & h_3 & b_\perp & \ldots &0 \\
0 & b_\perp  &\ddots  &  \\
\vdots &&&& \vdots \\
0 & \ldots & & h_{n_z-2} & b_\perp\\ 
0 & \ldots && b_\perp & h_{n_z-1} 
\end{array} \right ]
\left [ \begin{array}{c} Z_{:,2}^{(n)} \\ Z_{:,3}^{(n)} \\   \\\vdots \\  \\Z_{:,n_1-1}^{(n)} \end{array}\right ]
= \tilde \epsilon_n  \left [ \begin{array}{c} Z_{:,2}^{(n)} \\ Z_{:,3}^{(n)} \\   \\ \vdots \\   \\Z_{:,n_1-1}^{(n)} \end{array}\right ]
\end{equation}
where the subscript notation $Z^{(n)}_{:,j}$ refers to the entire $j$th column of the matrix ${\bf Z}^{(n)}$,
\begin{equation}
b_\perp = -\frac{\hbar^2}{2m^\ast \Delta^2} I_{n_x},
\end{equation}
with $I_{n_x}$ the $n_x\times n_x$ identity matrix, and
\begin{equation}
h_j = 
\left [ \begin{array}{cccccc} 
a_{1j} & b_x & 0 && \ldots &b_x \\
b_x & a_{2j} & b_x && \ldots &0 \\
\vdots &&&&& \vdots \\
0 & \ldots && b_x & a_{n_x-1, j} & b\\ 
b_x & \ldots &&& b_x & a_{n_x, j} 
\end{array} \right ]
\end{equation}
where 
\begin{eqnarray}
a_{ij} &=& 4 \frac{\hbar^2}{2m^\ast \Delta^2} - e\phi(x_i,z_j), \\
b_x &=& -\frac{\hbar^2}{2m^\ast \Delta^2}.
\end{eqnarray}
For a given potential $\phi(x,z)$, this can be diagonalized numerically to find the matrix of eigenvectors ${\bf Z}^{(n)}$ and eigenvalues $\tilde \epsilon_n$. 

To obtain the electron density for fixed 2D electron density $\ntd$, one first needs to obtain the chemical potential from
\begin{equation}
\ntd = \frac{2}{L_xL_y} \sum_{k_y, n} f( \epsilon_{n,k_y} )
\end{equation}
where the factor of 2 is for spin and $f(x)$ is the fermi-Dirac function.  Defining $\epsilon = \hbar^2 k_y^2/2m^\ast$, we have
\begin{eqnarray}
\ntd &=& \frac{2}{L_x}\sum_{n} \int \frac{dk_y}{2\pi} f(\tilde \epsilon_n + \epsilon) \nonumber \\  
&=& \sqrt{\frac 1\pi \left( \frac{2 m^\ast\Delta^2}{\hbar^2}\right ) k_B T } \nonumber \\
&&\times  \frac{1}{n_x\Delta^2} \sum_n \frac{1}{\sqrt{\pi}} \int_0^\infty \frac{y^{-1/2} dy} {e^{\beta(\tilde \epsilon_n -\mu) + y}+1} \nonumber \\ 
&=& \sqrt{\frac{2 m^\ast \Delta^2 k_BT}{\pi\hbar^2 }} \frac{1}{n_x\Delta^2} \sum_n {\cal F}_{-\frac 12}\left( \frac{\tilde \epsilon_n -\mu}{k_BT} \right ), \nonumber \\
\label{eq:n2D}
\end{eqnarray}
where ${\cal F}_{-\frac 12}$ is the Fermi-Dirac integral of order $-\frac 12$.

The three-dimensional electron density, which is required to calculate the electric potential from Poisson's equation, is then
\begin{equation}
n_e(x_i,z_j) =  \sqrt{\frac{2 m^\ast  k_BT}{\pi\hbar^2 \Delta^4}} \sum_n |Z^{(n)}_{ij}|^2 {\cal F}_{-\frac 12 }\left( \frac{\tilde \epsilon_n -\mu}{k_BT} \right ).
\end{equation}


%

\end{document}